\documentclass[ amsmath, amssymb, aps, prd, twocolumn, lengthcheck,
sort&compress, showpacs, superscriptaddress, nofootinbib ]{revtex4-1}

\usepackage[utf8x]{inputenc}
\usepackage{amssymb}
\usepackage{amsmath}
\usepackage{nicefrac}
\usepackage{multirow}
\usepackage{bbold}
\usepackage{longtable}

\usepackage{graphicx}% Include figure files
\usepackage{array}
\usepackage{dcolumn}% Align table columns on decimal point
\usepackage{mathrsfs}
\usepackage{comment}
\usepackage{amsbsy}
\usepackage{units}

\newcommand{\Slash}[1]{\ooalign{\hfil/\hfil\crcr$#1$}}

\newcommand{\Eq}[1]{(\ref{#1})}

\usepackage[usenames]{color}
\usepackage[colorlinks=true,linkcolor=blue,citecolor=blue,urlcolor=blue]{hyperref}

\definecolor{lred}{rgb}{1,0.90,0.7}

\begin{document}

\title{Delta and Omega electromagnetic form factors in a three-body covariant
Bethe-Salpeter approach}
\author{Helios \surname{Sanchis-Alepuz}}
\email{helios.sanchis-alepuz@theo.physik.uni-giessen.de}
\affiliation{Institute of Theoretical Physics, Justus-Liebig University of Gie\ss en, Heinrich-Buff-Ring 16, 35392, Gie\ss en, Germany}
\author{Richard \surname{Williams}}
\affiliation{Institute of Physics, University of Graz, Universit\"atsplatz 5, 8010, Graz, Austria}
\author{Reinhard \surname{Alkofer}}
\affiliation{Institute of Physics, University of Graz, Universit\"atsplatz 5, 8010, Graz, Austria}

\begin{abstract}
The electromagnetic form factors of the $\Delta$ and $\Omega$ baryons are calculated in the framework of Poincar\'e-covariant bound-state equations.
The quark-quark interaction is truncated to a single dressed-gluon exchange where for the dressings we use two different models and compare the results.
The calculation predicts an oblate shape for the electric charge distribution and a prolate shape for the magnetic dipole distribution.
We also identify the necessity of including pion-cloud corrections at low photon-momentum transfer.
\end{abstract}


\maketitle
\date{\today}

%%%%%%%%%%%%%%%%%%%%%%%%%%%%%%%%%%%%%%%%%%%%%%%%%%%%%%%%%%%%%%%%%%%%%
%                        S E C T I O N                              %
%%%%%%%%%%%%%%%%%%%%%%%%%%%%%%%%%%%%%%%%%%%%%%%%%%%%%%%%%%%%%%%%%%%%%

\section{Introduction}\label{sec:introduction}

The spatial distribution of hadrons' extensive properties, such as mass or
electric charge, is of especial relevance in the understanding of low-energy QCD
dynamics, since they probe the details of the quark-quark and gluon-quark
interactions.

The electromagnetic properties of the proton have been widely studied experimentally,
providing a good picture of its internal structure. This is not the case,
however, for the lightest baryonic resonance, the $\Delta(1232)$. Its short
lifetime makes the study of its properties difficult and only the magnetic
moments of $\Delta^{++}$
\cite{Nefkens:1977eb,Heller:1986gn,Wittman:1987kb,Lin:1991tx,Lin:1991qk,
Bosshard:1990ys,Bosshard:1991zp,LopezCastro:2000cv,LopezCastro:2000ep,
Beringer:1900zz} and $\Delta^{+}$ \cite{Beringer:1900zz,Kotulla:2002cg} are
known, albeit with large errors. An indirect estimation of the $\Delta^{+}$
electric quadrupole moment from the $\gamma N\rightarrow\Delta$ transition
quadrupole moment was given in \cite{Blanpied:2001ae}. The $\Omega(1672)$ decays
weakly, instead, and this has allowed for a precise measurement of its magnetic 
dipole moment~\cite{Beringer:1900zz}.

For finite values of the photon momentum the only information available comes
from lattice QCD 
calculations~\cite{Alexandrou:2009hs,Alexandrou:2009nj,Alexandrou:2010jv,Aubin:2008qp,Boinepalli:2009sq}. 
Although constantly improving, these calculations suffer
from the usual problem of not yet being able to work at the physical pion mass.
Moreover, the limit of vanishing photon momentum is unreachable for technical
reasons. The calculation of the electromagnetic properties of the Delta and
Omega baryons has also been tackled from a number of constituent quark models
\cite{Ramalho:2008dc,Ramalho:2009vc,Ramalho:2010xj,Ramalho:2010rr}, chiral
quark-soliton model \cite{Ledwig:2008es}, chiral perturbation theory
\cite{Geng:2009ys,Ledwig:2011cx} and QCD sum rules \cite{Azizi:2008tx}.

In this paper, we investigate the electromagnetic properties of the Delta
and Omega baryon in the framework of covariant Bethe-Salpeter equations.
In section~\ref{sec:current} we introduce the general formalism of 
Bethe-Salpeter equations (BSE) and Dyson-Schwinger equations (DSE). This is followed by a presentation
of the truncation used in section~\ref{sec:truncation}. In section~\ref{sec:results}
the results of our calculation are discussed. Finally, we conclude in
section~\ref{sec:summary}.

%%%%%%%%%%%%%%%%%%%%%%%%%%%%%%%%%%%%%%%%%%%%%%%%%%%%%%%%%%%%%%%%%%%%%
%                        S E C T I O N                              %
%%%%%%%%%%%%%%%%%%%%%%%%%%%%%%%%%%%%%%%%%%%%%%%%%%%%%%%%%%%%%%%%%%%%%

\section{Baryon Bethe-Salpeter equation and coupling to an external
field}\label{sec:current} 

\begin{figure}[ht!]
 \begin{center}
  \includegraphics[width=0.4\textwidth,clip]{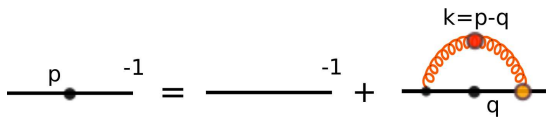}
 \end{center}
 \caption{Diagrammatic representation of the quark Dyson-Schwinger equation
\Eq{eq:quarkDSE}. Blobs represent fully dressed propagators or
vertices.}\label{fig:quarkDSE}
\end{figure}

\begin{figure*}[ht]
 \begin{center}
  \includegraphics[width=0.8\textwidth,clip]{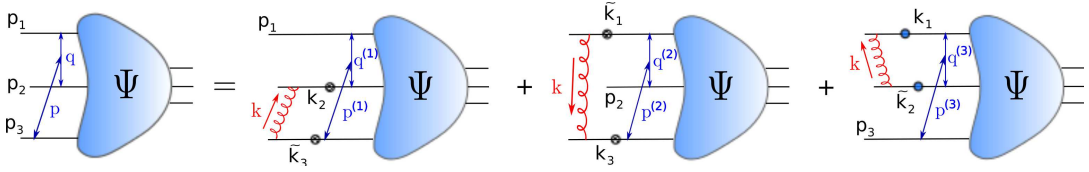}
 \end{center}
 \caption{Diagrammatic representation of the covariant Faddeev equation in the
Rainbow-Ladder truncation \Eq{eq:faddeev_eq}.}\label{fig:FaddeevRLeq}
\end{figure*}

The evolution of a three-quark system in quantum field theory is described
through the six-quark Green's function $G^{(3)}(p_1,p_2,p_3)$ (in momentum
space) or, equivalently, its amputated version the scattering matrix
$T^{(3)}(p_1,p_2,p_3)$. This function can be obtained by solving a Dyson
equation\footnote{For simplicity, we employ a compact matrix notation in which discrete/continuous
variables are implicitly summed/integrated over.}
\begin{equation}\label{eq:DysonEq}
 T=-iK-iKG_0T~,
\end{equation}
or, equivalently,
\begin{align}
 T^{-1}&=iK^{-1}-G_0 \,\,,\label{eq:DysonEq-b}\\
\mathbb{1}&=iTK^{-1}-TG_0\,\,,\label{eq:DysonEq-c}
\end{align}
where $G_0$ is the disconnected product of three full quark propagators and $-iK$ is
the three-quark interaction kernel. The latter includes three- and two-particle
irreducible interactions
\begin{equation}\label{eq:def_irred_kernels}
 K\equiv \widetilde{K}^{(3)}+\sum_{a=1}^3 \widetilde{K}^{(2)}_aS^{-1}_a\,\,,
\end{equation}
with $a$ denoting the spectator quark (see e.g. Fig.~\ref{fig:FaddeevRLeq}).
The full quark propagator $S$ is obtained by solving the quark propagator DSE
(see Fig.~\ref{fig:quarkDSE})
\begin{equation}\label{eq:quarkDSE}
 S^{-1}(p)=S^{-1}_0(p)+Z_{1f}\int_q \gamma^\mu
D_{\mu\nu}(p-q)\Gamma^\nu_{gqq}(p,q)S(q)\,\,,
\end{equation}
where $S_0$ is the (renormalized) bare propagator
\begin{equation}\label{eq:bare_prop}
 S^{-1}_0(p)=Z_2\left(i\Slash{p}+m\right)\,\,,
\end{equation}
 $\Gamma^\nu_{gqq}$ is the full quark-gluon vertex and $D^{\mu\nu}$ is the full
gluon propagator and $Z_{1f}$ and $Z_2$ are renormalization constants. 
In the Landau gauge the gluon propagator takes the form
\begin{equation}
 D^{\mu\nu}(k)=T^{\mu\nu}(k)\frac{Z(k^2)}{k^2}~,
\end{equation}
with $Z(k^2)$ a dressing function to be determined and $T^{\mu\nu}$ the
transverse projector
\begin{equation}\label{eq:def_transverse_proj}
 T_{\mu\nu}(k)=\delta_{\mu\nu}-\frac{k_\mu k_\nu}{k^2}\,\,.
\end{equation}
The bare quark mass $m$ in \Eq{eq:bare_prop} must be provided as a parameter.

When the three-quark system forms a bound state, the Green's function $T^{(3)}$
develops a pole at $P^2=-M^2$, with $P$ the total quark momentum
\begin{equation}
 P=p_1+p_2+p_3\,\,,
\end{equation}
and $p_i$ the quark momenta, with $M$ the bound-state mass. At the bound state pole one defines
\begin{equation}\label{eq:def_BSEamplitudes}
 T^{(3)}\sim\mathcal{N}\frac{\Psi\bar{\Psi}}{P^2+M^2}~,
\end{equation}
where $\mathcal{N}$ is a normalization factor which, in the case of
spin-$\nicefrac{3}{2}$ particles is $2M$. The function $\Psi$ is the bound-state
Bethe-Salpeter amplitude and $\bar{\Psi}$ its charge conjugate. They are
expressed as tensor products of flavor, color and spin parts which describe a
baryon in terms of its constituent quarks. For spin-$\nicefrac{3}{2}$ baryons
the spin part is itself a mixed tensor with four Dirac indices and one Lorentz
index \cite{SanchisAlepuz:2010in,SanchisAlepuz:2011jn,SanchisAlepuz:2012vb}.
Substituting \Eq{eq:def_BSEamplitudes} in \Eq{eq:DysonEq-b} or in
\Eq{eq:DysonEq-c}, and keeping only the singular terms, we arrive at the
Bethe-Salpeter equation for the three-quark bound state
\begin{equation}\label{eq:compactBSE}
 \Psi=-iKG_0\Psi \,\,,
\end{equation}
or
\begin{equation}\label{eq:compactBSE-conj}
 i\bar{\Psi}K^{-1}=\bar{\Psi}G_0 \,\,.
\end{equation}

A systematic procedure to couple an external gauge field to the constituents of
a three-particle system described by integral equations is the so-called gauging
of the equations, introduced in \cite{Haberzettl:1997jg,Kvinikhidze:1998xn,Kvinikhidze:1999xp,Oettel:1999gc}. It
ensures that the resulting equations are gauge invariant and that there is no
over-counting of diagrams. For our purposes it suffices to say that the gauging
of equations acts as a derivative on the integral equation. That is,
\Eq{eq:DysonEq} becomes
\begin{flalign}
 T^{\mu}=&~-iK^\mu-iK^\mu G_0T\nonumber\\
             &~-iKG^{\mu}_0T-iKG_0T^{\mu}\,\,,
\end{flalign}
where the superindex $\mu$ denotes a \textit{gauged} function (that is, coupled
to the gauge field). This equation can be rewritten, using \Eq{eq:DysonEq}, as
\begin{flalign}\label{eq:Tmu_equation}
 T^{\mu}=&~(1+iKG_0)^{-1}\times\nonumber\\
            &\left(-iK^\mu-iK^\mu
G_0T-iKG^{\mu}_0T\right)\nonumber\\
            =&~T\left(iK^{-1}K^\mu K^{-1}+G^{\mu}_0\right)T~.
\end{flalign}
To have an expression for $K^\mu$ one needs to specify the interaction kernel.
In the next section we shall obtain the gauged kernel in the case of the
Rainbow-Ladder truncation. The gauged quark propagator allows the introduction of the
proper vertex $\Gamma^\mu$ through the definition
\begin{equation}\label{eq:definition_qphvertex}
 S^\mu=S\Gamma^\mu S\,\,,
\end{equation}
which, in the case that concerns us in this paper, represents the fully dressed 
quark-photon vertex.

The bound-state electromagnetic current $J^\mu$ can be introduced at the
bound-state pole by
\begin{flalign}
 T^{(3),\mu}\sim\mathcal{N}_i\mathcal{N}_f\frac{\Psi_f}{P_f^2+M_f^2}J^\mu\frac{\bar{\Psi}_i}{P_i^2+M_i^2}
\,\,.
\end{flalign}
Substituting in \Eq{eq:Tmu_equation} and using \Eq{eq:compactBSE} and \Eq{eq:compactBSE-conj}
we arrive at
\begin{flalign}\label{eq:current_general}
 J^\mu=&~\bar{\Psi}_f\left(-iG_0K^\mu G_0+G^{\mu}_0\right)\Psi_i\,\,.
\end{flalign}
The electromagnetic form factors are calculated via the identification of
\Eq{eq:current_general} with the expression of the current imposed by
symmetry principles (see Appendix \ref{sec:appendix_current}).

%%%%%%%%%%%%%%%%%%%%%%%%%%%%%%%%%%%%%%%%%%%%%%%%%%%%%%%%%%%%%%%%%%%%%
%                        S E C T I O N                              %
%%%%%%%%%%%%%%%%%%%%%%%%%%%%%%%%%%%%%%%%%%%%%%%%%%%%%%%%%%%%%%%%%%%%%

\section{Rainbow-Ladder truncation}\label{sec:truncation}

To solve \Eq{eq:compactBSE} one needs to specify the interaction kernel
$K$. An exact expression for this kernel is in general not available and one has
to resort to some truncation scheme. The simplest consistent scheme is known as
Rainbow-Ladder (RL) truncation. This truncation preserves the axial-vector
Ward-Takahashi identity, which relates the quark-antiquark interaction kernel
and the quark-gluon vertex in the quark DSE \cite{Munczek:1994zz,Bender:1996bb}.
In the meson sector this identity ensures that pions are the Goldstone bosons of
spontaneous chiral symmetry breaking \cite{Maris:1997hd}. The RL truncation
reduces the quark-antiquark kernel to a single dressed-gluon exchange. The full
quark-gluon vertex is projected onto the tree-level Lorentz structure
$\gamma^\mu$ and the non-perturbative dressing is restricted to depend on the
gluon momentum only and has to be modelled. It is customary to include this
dressing and the gluon propagator dressing $Z(k^2)$ in a single effective
interaction $\alpha(k^2)$.

\subsection{Three-quark bound state equations}

Interactions in the baryon sector are not restricted, in principle, by the
axial-vector Ward-Takahashi identity. However, we adopt here for the quark-quark
interaction kernel the same truncation scheme and neglect the three-particle
irreducible interactions. The three-body BSE \Eq{eq:compactBSE} in the RL
truncation, which is also known as covariant Faddeev equation (and,
correspondingly, the Bethe-Salpeter amplitudes $\Psi$ are called Faddeev
amplitudes), reads (see Fig.~{\ref{fig:FaddeevRLeq})
\begin{widetext}
\begin{flalign}\label{eq:faddeev_eq}
\Psi_{\alpha\beta\gamma\mathcal{I}}(p,q,P) ={}&\int_k  \left[
\widetilde{K}_{\beta\beta'\gamma\gamma'}(k)~S_{\beta'\beta''}(k_2)
S_{\gamma'\gamma''}(\tilde{k}_3)~
\Psi_{\alpha\beta''\gamma''\mathcal{I}}(p^{(1)},q^{(1)},P)\right.\nonumber\\
&\quad \left.
+\widetilde{K}_{\alpha\alpha'\gamma\gamma'}(-k)~S_{\gamma'\gamma''}(k_3)
S_{\alpha'\alpha''} (\tilde{k}_1)~
\Psi_{\alpha''\beta\gamma''\mathcal{I}}(p^{(2)},q^{(2)},P)
\right. \nonumber\\
&\quad  \left. +
\widetilde{K}_{\alpha\alpha'\beta\beta'}(k)~S_{\alpha'\alpha''}(k_1)
S_{\beta'\beta''}(\tilde{k}_2)~
\Psi_{\alpha''\beta''\gamma\mathcal{I}}(p^{(3)},q^{(3)},P)\right]\,\,,
\end{flalign}
\end{widetext}
where we have absorbed the $-i$ factor into the definition of $\widetilde{K}$,
so that it is defined as
\begin{equation}\label{eq:RLkernel}
	\widetilde{K}_{\alpha\alpha'\beta\beta'}(k)= -4\pi C~Z_2^2
~\frac{\alpha_{\textrm{eff}}(k^2)}{k^2}~
	T_{\mu\nu}(k)~\gamma^\mu_{\alpha\alpha'}  \gamma^\nu_{\beta\beta'}\,\,,
\end{equation}
with $Z_2$ the quark renormalization constant. We have
used the generic index $\mathcal{I}$ to refer to the bound state (as opposed to the first
three Greek indices in the Faddeev amplitude which denote the valence quarks);
for the case of a spin-$\nicefrac{3}{2}$ baryon $\mathcal{I}$ consists of a
Dirac and a Lorentz index. In \Eq{eq:faddeev_eq}, the flavor part of the Faddeev
amplitudes has been factored out because the interaction kernel is
flavor-independent and the factor $C=-2/3$~ stems from the traces of the color
structures.
The Faddeev amplitudes depend on the quark momenta $p_1$, $p_2$ and $p_3$, but
this dependence can be reexpressed in terms of the total momentum $P$ and two
relative momenta $p$ and $q$
\begin{align}\label{eq:defpq}
        p &= (1-\zeta)\,p_3 - \zeta (p_1+p_2)\,, &  p_1 &=  -q -\dfrac{p}{2} +
\dfrac{1-\zeta}{2} P\,, \nonumber\\
        q &= \dfrac{p_2-p_1}{2}\,,         &  p_2 &=   q -\dfrac{p}{2} +
\dfrac{1-\zeta}{2} P\,, \nonumber\\
        P &= p_1+p_2+p_3\,,                &  p_3 &=   p + \zeta  P\,\,,\nonumber\\
\end{align}
with $\zeta$ a free momentum partitioning parameter, which we choose
$\zeta=1/3$ for numerical convenience. The internal quark propagators depend on 
the internal quark momenta
$k_i=p_i-k$ and $\tilde{k}_i=p_i+k$, with $k$ the gluon momentum. The internal
relative momenta, for each of the three terms in the Faddeev equation, are
\begin{align}\label{internal-relative-momenta}
p^{(1)} &= p+k,& p^{(2)} &= p-k,& p^{(3)} &= p,\nonumber\\
q^{(1)} &= q-k/2,& q^{(2)} &= q-k/2, & q^{(3)} &= q+k\,\,.\nonumber\\
\end{align}
The quark DSE in the RL truncation reduces to 
\begin{equation}\label{eq:quarkDSERL}
 S^{-1}_{\alpha'\beta'}(p)=S_{0,\alpha'\beta'}(p)+\int_q
\widetilde{K}_{\alpha\alpha'\beta\beta'}(k)S_{\alpha'\beta'}(q)\,\,,
\end{equation}
where now in $\widetilde{K}$, see \Eq{eq:RLkernel}, we have $C=4/3$ and $k=p-q$.

\subsection{Bound state electromagnetic current and quark-photon vertex}

The expression for the current \Eq{eq:current_general} simplifies
considerably in the RL truncation. Since $\widetilde{K}^{(3)}$ is absent and
$\widetilde{K}^{(2)}$ is reduced to the exchange of a neutral particle, the
photon can only couple to the quark propagator through the term $S^{-1}$ in
\Eq{eq:def_irred_kernels}. Defining $\mathbb{1}^\mu=(SS^{-1})^\mu=0$ we
obtain
\begin{equation}
 \left(S^{-1}\right)^\mu=-S^{-1}S^\mu S^{-1}=-\Gamma^\mu\,\,,
\end{equation}
where we used \Eq{eq:definition_qphvertex}. Thus \Eq{eq:current_general} becomes
({\it cf.} Fig.~\ref{fig:FFeq_RL})
\begin{figure*}[hbtp]
 \begin{center}
  \includegraphics[width=0.7\textwidth,clip]{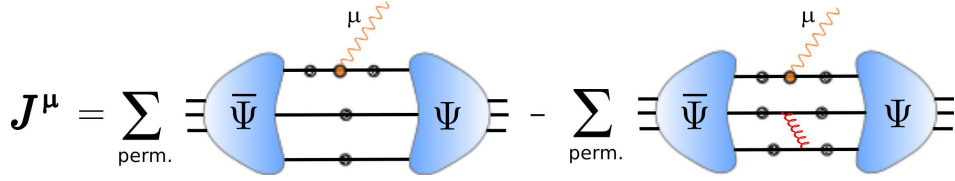}
 \end{center}
 \caption{Diagrammatic representation of the current
\Eq{eq:current_general} in
the Rainbow-Ladder truncation, see Eq.\ \Eq{eq:FFeqRL}.}\label{fig:FFeq_RL}
\end{figure*}
\begin{widetext}
\begin{flalign}\label{eq:FFeqRL}
 J_{\mathcal{I}'\mathcal{I}}^\mu=\int_p\int_q\bar{\Psi}_{\beta'\alpha'\mathcal{I
}'\gamma'}(p_f^{\{1\}},q^{\{1\}}_f,P_f)\left[\left(S(p_1^f)\Gamma^\mu(p_1,
Q)S(p_1^i)\right)_{\alpha'\alpha}S_{\beta'\beta}(p_2)S_{\gamma'\gamma}
(p_3)\right]\times\nonumber\\
~~~~~~~~~~~~~~~~~~~~~~~~~~~~~~\left(\Psi_{\alpha\beta\gamma\mathcal{I}}(p^{\{1\}
}_i,q^{\{1\}}_i,P_i)-\Psi^{\{1\}}_{\alpha\beta\gamma\mathcal{I}}(p^{\{1\}}_i,q^{
\{1\}}_i,P_i)\right)\nonumber\\
+\int_p\int_q\bar{\Psi}_{\beta'\alpha'\mathcal{I}'\gamma'}(p^{\{2\}}_f,q^{\{2\}}
_f,P_f)\left[S_{\alpha'\alpha}(p_1)\left(S(p_2^f)\Gamma^\mu(p_2,
Q)S(p_2^i)\right)_{\beta'\beta}S_{\gamma'\gamma}(p_3)\right]\times\nonumber\\
~~~~~~~~~~~~~~~~~~~~~~~~~~~~~~\left(\Psi_{\alpha\beta\gamma\mathcal{I}}(p^{\{2\}
}_i,q^{\{2\}}_i,P_i)-\Psi^{\{2\}}_{\alpha\beta\gamma\mathcal{I}}(p^{\{2\}}_i,q^{
\{2\}}_i,P_i)\right)\nonumber\\
+\int_p\int_q\bar{\Psi}_{\beta'\alpha'\mathcal{I}'\gamma'}(p^{\{3\}}_f,q^{\{3\}}
_f,P_f)\left[S_{\alpha'\alpha}(p_1)S_{\beta'\beta}
(p_2)\left(S(p_3^f)\Gamma^\mu(p_3,Q)S(p_3^i)\right)_{\gamma'\gamma}\right]
\times\nonumber\\
~~~~~~~~~~~~~~~~~~~~~~~~~~~~~~\left(\Psi_{\alpha\beta\gamma\mathcal{I}}(p^{\{3\}
}_i,q^{\{3\}}_i,P_i)-\Psi^{\{3\}}_{\alpha\beta\gamma\mathcal{I}}(p^{\{3\}}_i,q^{
\{3\}}_i,P_i)\right)~,
\end{flalign}
%\end{widetext}
%
where we defined
%
%\begin{widetext}
\begin{flalign}\label{eq:third_diagram}
 \Psi^{\{1\}}_{\alpha\beta\gamma\mathcal{I}}=
\int_k  \widetilde{K}_{\beta\beta'\gamma\gamma'}(k)~S_{\beta'\beta''}(p_2-k)
S_{\gamma'\gamma''}(p_3+k)~
\Psi_{\alpha\beta''\gamma''\mathcal{I}}(p+k,q-k/2,P)\,\,,
\end{flalign}
\end{widetext}
as a result of the first term in the Faddeev equation \Eq{eq:faddeev_eq}
and in a similar fashion we define $\Psi^{\{2\}}$ and $\Psi^{\{3\}}$.  The \textit{injected} momentum $Q$ is introduced via
the final and initial momenta of the interacting quark $\kappa$
\begin{equation}
 p_\kappa^{\nicefrac{f}{i}}=p_\kappa\pm\frac{Q}{2}~.
\end{equation}
The relative momenta in the respective terms of \Eq{eq:FFeqRL} are, using the
definitions in \Eq{eq:defpq},
\begin{align}\label{eq:relative_momenta_withQ}
 p_{\nicefrac{f}{i}}^{\{1\}}&=p\mp\zeta\frac{Q}{2}~,&q_{\nicefrac{f}{i}}^{\{1\}}
&=q\mp\frac{Q}{4}~, \nonumber\\
 p_{\nicefrac{f}{i}}^{\{2\}}&=p\mp\zeta\frac{Q}{2}~,&q_{\nicefrac{f}{i}}^{\{2\}}
&=q\pm\frac{Q}{4}~, \\
 p_{\nicefrac{f}{i}}^{\{3\}}&=p\pm(1-\zeta)\frac{Q}{2}~,&q_{\nicefrac{f}{i}}^{\{
3\}}&=q~, \nonumber
\end{align}
and since the initial and final states are on-shell, the total momenta are
constrained to be $P_i^2=P_f^2=-M^2$, with $M$ the mass of the bound-state. 
As is the case for the Faddeev equation, the three terms in \Eq{eq:FFeqRL} are formally the
same when the momentum partitioning parameter is chosen to be $\zeta=1/3$.

The quark-photon vertex $\Gamma^\mu$ can naturally be incorporated in the
framework of covariant bound-state equations by calculating it from an
inhomogenous Bethe-Salpeter equation
\begin{flalign}
 \Gamma^\mu(p,Q)&=iZ_2\gamma^\mu\nonumber\\                
&+\int_kK_{q\bar{q}}\left(S(k+Q/2)\Gamma^\mu(k,Q)S(k-Q/2)\right)~,
\end{flalign}
and using for $K_{q\bar{q}}$ the RL kernel \Eq{eq:RLkernel} with $C=4/3$ and
for the quark propagator $S$ the solutions of the RL-truncated quark DSE
\Eq{eq:quarkDSERL}. We calculate this in the appropriate moving frame
following Ref.~\cite{Bhagwat:2006pu}.

%%%%%%%%%%%%%%%%%%%%%%%%%%%%%%%%%%%%%%%%%%%%%%%%%%%%%%%%%%%%%%%%%%%%%
%                      S U B S E C T I O N                          %
%%%%%%%%%%%%%%%%%%%%%%%%%%%%%%%%%%%%%%%%%%%%%%%%%%%%%%%%%%%%%%%%%%%%%

\subsection{Effective interactions}

The appearance of the effective interaction in \Eq{eq:RLkernel} will
\textit{a priori} introduce a model dependence on our results. In fact, this is
the only model input of the approach. To assess how strong is this dependence
and to identify the possible model-independent features, we use two different
models for the effective interaction in our calculations.

The first model we use is known as the Maris-Tandy model~\cite{Maris:1997tm,Maris:1999nt} 
and has dominated hadron studies within Rainbow-Ladder. This dominance is well-earned 
since this ansatz performs very
well when it comes to the purely phenomenological calculation of ground-state
meson and baryon properties. However, this model has no clear connection to QCD
in the intermediate- and low-momentum regime and is, therefore, not entirely
satisfactory to gain understanding of the formation of hadronic bound-states in
QCD. On the other hand, with the rapid improvement in our knowledge of QCD
Green's
functions from both lattice and functional approaches, it is possible to
define different effective interactions which, presumably, capture more
faithfully some of QCD's features. Based on this, an effective interaction has
been proposed in Ref.~\cite{Alkofer:2008et}.

Note that the fact that an effective interaction captures more features of QCD
does not necessarily mean that it will perform better phenomenologically. This
is because the interaction is used within a given truncation scheme and,
therefore, if one wants to reproduce hadron properties the model has to be tuned
to account for the effect of the missing contributions. In particular, it has
been shown in~\cite{Alkofer:2008tt} that dynamical quark-mass generation is
accompanied by the appearance of scalar components in the quark-gluon vertex.
An application of this beyond Rainbow-Ladder has been pursued 
in Refs.~\cite{Fischer:2009jm,Williams:2009wx}
with a non-diagrammatic means provided in Ref.~\cite{Chang:2009zb}. 
In addition, unquenching effects in the form of a pion back coupling to the
quark propagator and two-body kernel have been investigated in Refs.~\cite{Fischer:2007ze,Fischer:2008wy}.
However, 
none of these methods have yet been extended to the covariant three-body problem
presented here and so we restrict ourselves to Rainbow-Ladder. Since we
lose many components of the quark-gluon vertex we therefore construct an
effective interaction that attempts to mimic their contribution.

In this respect, both models described below are designed to correctly reproduce 
dynamical chiral-symmetry breaking as well as pion properties at the physical
$u/d$ mass. This means that they account for missing effects in the bound-state 
pseudoscalar meson sector and at this quark mass. As a consequence,
both interactions have similar strength at the intermediate momentum region
$\sim0.5~-~1~$GeV (see Fig.~\ref{fig:Int_comparison}).  These two interactions have previously
been compared in Ref.~\cite{SanchisAlepuz:2011aa}.

\begin{figure}[b]
 \begin{center}
  \includegraphics[width=0.4\textwidth,clip]{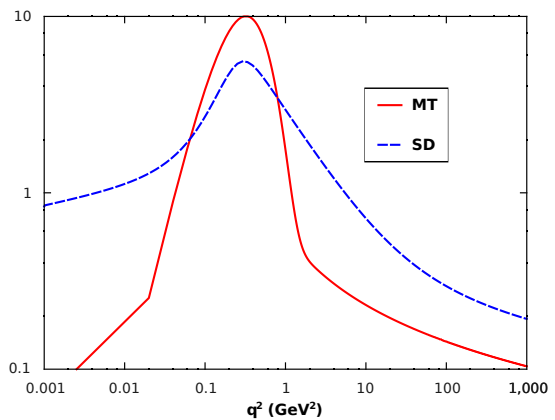}
 \end{center}
 \caption{Comparison of the Maris-Tandy (MT) \Eq{eq:MTmodel}, with $\eta=1.8$, 
and the soft-divergence (SD)
\Eq{eq:AFWmodel} effective interactions.}\label{fig:Int_comparison}
\end{figure}

%%%%%%%%%%%%%%%%%%%%%%%%%%%%%%%%%%%%%%%%%%%%%%%%%%%%%%%%%%%%%%%%%%%%%
%                      S U B S E C T I O N                          %
%%%%%%%%%%%%%%%%%%%%%%%%%%%%%%%%%%%%%%%%%%%%%%%%%%%%%%%%%%%%%%%%%%%%%

\subsubsection{Maris-Tandy model}

In the Maris-Tandy (MT) model \cite{Maris:1997tm,Maris:1999nt} the effective
running
coupling is given by
\begin{flalign}\label{eq:MTmodel}
\alpha_{\textrm{eff}}(q^2) {}=&
 \pi\eta^7\left(\frac{q^2}{\Lambda^2}\right)^2
e^{-\eta^2\frac{q^2}{\Lambda^2}}\nonumber\\ &+{}\frac{2\pi\gamma_m
\big(1-e^{-q^2/\Lambda_{t}^2}\big)}{\textnormal{ln}[e^2-1+(1+q^2/\Lambda_{QCD}
^2)^2]}\,, 
\end{flalign}
which reproduces the one-loop QCD behavior in the UV and features a Gaussian
distribution in the intermediate momentum region (see Fig.~\ref{fig:Int_comparison}) 
that provides dynamical chiral symmetry breaking. The
scale $\Lambda_t=1$~GeV is introduced for technical reasons and has no impact on
the results. Therefore, the interaction strength is characterized by an energy
scale $\Lambda$, fixed to $\Lambda=0.74$ GeV to reproduce correctly the pion
decay constant from the RL-truncated meson-BSE. The dimensionless parameter
$\eta$ controls the width of the interaction. 
For the anomalous dimension we use $\gamma_m=12/(11N_C-2N_f)=12/25$,
corresponding to $N_f=4$ flavors and $N_c=3$ colors. For the QCD scale
$\Lambda_{QCD}=0.234$ GeV. Many ground-state hadron observables have been found
to be almost insensitive to the value of $\eta$ around $\eta=1.8$ (see, e.g.
\cite{Krassnigg:2009zh,Eichmann:2011vu,Nicmorus:2010mc}). This has been used as
an argument in favor of the model independence of Rainbow-Ladder results.
Instead of pursuing this line of research, we prefer to introduce a new,
non-related model to evaluate the validity of those assertions.

Note that in the numerical resolution of the quark DSE we employ the
Pauli-Villars regularization method of the integrals, with a mass scale of
$200$~GeV.
Moreover, for this model, we fit the quark masses, at the renormalization scale
$\mu=19$~GeV, to be $3.7$, $85.2$, $869$ and $3750$ MeV for the
$u/d$, $s$, $c$, and $b$ quarks, respectively.

%%%%%%%%%%%%%%%%%%%%%%%%%%%%%%%%%%%%%%%%%%%%%%%%%%%%%%%%%%%%%%%%%%%%%
%                      S U B S E C T I O N                          %
%%%%%%%%%%%%%%%%%%%%%%%%%%%%%%%%%%%%%%%%%%%%%%%%%%%%%%%%%%%%%%%%%%%%%

\subsubsection{Soft-divergence model}

The model of Ref.~\cite{Alkofer:2008et}, called soft-divergence or SD
model from here on, is motivated by
the desire to account
for the $U_A(1)$-anomaly by the Kogut-Susskind
mechanism \cite{Kogut:1973ab,vonSmekal:1997dq}. The effective coupling is
constructed as the product of the gluon dressing
\cite{Alkofer:2003jk,Alkofer:2003jj}
 and a model for
the non-perturbative behavior of the quark-gluon
vertex \cite{Alkofer:2008tt},
\begin{flalign}\label{eq:AFWmodel}
  \alpha_{\textrm{eff}}(q^2) {}=&  \mathcal{C} \left(\frac{x}{1+x}\right)^{2\kappa}
  \left(\frac{y}{1+y}\right)^{-\kappa-1/2}\nonumber\\
  &\times{}\left( \frac{\alpha_0+a_{UV}\,x}{1+x} \right)^{-\gamma_0}
  \left( \lambda +\frac{a_{UV}\,x}{1+x} \right)^{-2\delta_0}\,\,.
\end{flalign}
The four terms in parentheses are: the IR scaling of the gluon
propagator; IR scaling of the quark-gluon vertex; logarithmic running of
the gluon propagator; and the logarithmic running of the quark-gluon
vertex. Additionally, the last two are constructed to interpolate between the IR
and UV
behavior. The remaining terms are defined as
\begin{equation}
    \lambda = \frac{\lambda_S}{1+y} + \frac{\lambda_B \,y}{1+(y-1)^2}\,, \quad
    a_{UV}= \pi  \gamma_m \left( \frac{1}{\ln{z}}-\frac{1}{z-1} \right), \quad
\end{equation}
where
\begin{flalign}
       x &= q^2/\Lambda_{YM}^2\,, \\
       y &= q^2/\Lambda_{IR}^2\,, \\
       z &= q^2/\Lambda_{MOM}^2\,,
\end{flalign}
and $\alpha_0=8.915/N_C$. Here, $\Lambda_{YM}= 0.71$ GeV is the
dynamically generated Yang-Mills scale, while $\Lambda_{MOM}\simeq 0.5$ GeV
corresponds
to the one-loop perturbative running. The IR scaling exponent is
$\kappa=0.595353$, and the one-loop anomalous dimensions are
related via $1+\gamma_0 = -2\delta_0 = \frac{3}{8}\,N_C \,\gamma_m$, with
$\gamma_m=12/(11N_C-2N_f)$.
We choose $N_f=5$ active quark flavors at the
renormalization point $\mu=19$ GeV. The constant
$\mathcal{C}=0.968$ is chosen such that $\alpha_{\textrm{eff}}$ runs appropriately in
the UV. Finally, $\Lambda_{IR}=0.42$ GeV, $\lambda_S=6.25$, and
$\lambda_{B}=21.83$
determine the IR properties of the quark-gluon vertex and are fitted such that
the properties of $\pi$, $K$ and $\rho$
mesons are all reasonably well reproduced.  The quark
masses at $\mu=19$~GeV are $2.76$, $55.3$, $688$ and $3410$ MeV for the
$u/d$, $s$, $c$, and $b$ quarks, respectively.

\subsubsection{A remark on missing mesonic effects}

The MT and SD model both rely upon the phenomenology of dynamical chiral symmetry breaking in the light quark sector to determine their parameters. 
Therefore effects we might consider to be beyond RL are absorbed into the 
model parameterisation. In particular, since these are determined in the light-quark sector we
implicitly include those contributions due to interactions at the hadronic level.
Here the pion as the lightest hadron plays a special role in the dressing of baryons.
Amongst these contributions, non-perturbative pionic effects -- also sometimes called pion-cloud effects, see
\emph{e.g.} Ref.~\cite{Thomas:2007bc} and references therein -- are expected to have a
sizeable influence on hadron properties like the masses or the decay constants.
Consequently when fixing the model parameters in the light-quark sector, large parts of these
so-called pion-cloud contributions are ``parameterised'' in \emph{cf.}, the discussion
in Ref.~\cite{Eichmann:2008ae}.
According to Zweig's rule the meson-cloud around the triple-strange $\Omega$ will be
mostly constituted of kaons. Due to their higher mass as compared to pions, perturbative 
as well as non-perturbative mesonic effects are significantly smaller for the ground state 
properties of the $\Omega$ than for the  ground state properties of the $\Delta$. 
However, as we do not change the parameters of the model for the $\Omega$
we expect to find larger deviations from experimental values. This is because we actually 
then overestimate the  beyond RL effects; they look larger despite being actually smaller.
This should be kept in mind when comparing our results to lattice data and experimental 
observations. 

%%%%%%%%%%%%%%%%%%%%%%%%%%%%%%%%%%%%%%%%%%%%%%%%%%%%%%%%%%%%%%%%%%%%%
%                          S E C T I O N                            %
%%%%%%%%%%%%%%%%%%%%%%%%%%%%%%%%%%%%%%%%%%%%%%%%%%%%%%%%%%%%%%%%%%%%%

\section{Results}\label{sec:results}

We computed the electromagnetic current of the Delta and the Omega baryons and
extracted the corresponding form factors using \Eq{eq:expressionGs}. As
explained in previous sections, the interaction parameters and bare quark masses
are fitted to reproduce meson properties. In the baryon sector, therefore, there
are no further parameters to be fixed.

Of the four $\Delta(1232)$ isospin partners, we restrict the discussion to the $\Delta^+$ since, 
due to the assumption of isospin symmetry, the form factors of the remaining iso-partners can be 
obtained by multiplying with the corresponding baryon charge. This, in particular, implies that all 
$\Delta^0$ form factors are identically zero in our approach.

The solution of the
Faddeev equation, \Eq{eq:faddeev_eq}, and the subsequent calculation of the
electromagnetic current via \Eq{eq:FFeqRL} is a numerically complicated task,
chiefly as a consequence of the expansion of the Faddeev amplitudes in 128
Lorentz covariants and in a number of Chebyshev polynomials for the angular
dependence, which entails that one must solve for an equal number of
coefficients. Due to CPU-time and memory limitations, the number of quadrature
points used in the numerical integrations must be kept small. Moreover, the
presence of inverse powers of $Q$ in the equations for the extraction of the
form factors \Eq{eq:expressionGs} implies that, to obtain reliable results at
low $Q$ and even finite results in the limit $Q\rightarrow 0$, very delicate
cancelations among the many terms that contribute to the current must take
place. For these reasons, that limit is difficult to reach with our current resources, 
especially for the electric quadrupole, see also Ref.~\cite{SanchisAlepuz:2012ej}, and 
magnetic octupole form factors.

\begin{figure*}[hbtp]
 \begin{center}
  \includegraphics[width=0.9\textwidth,clip]{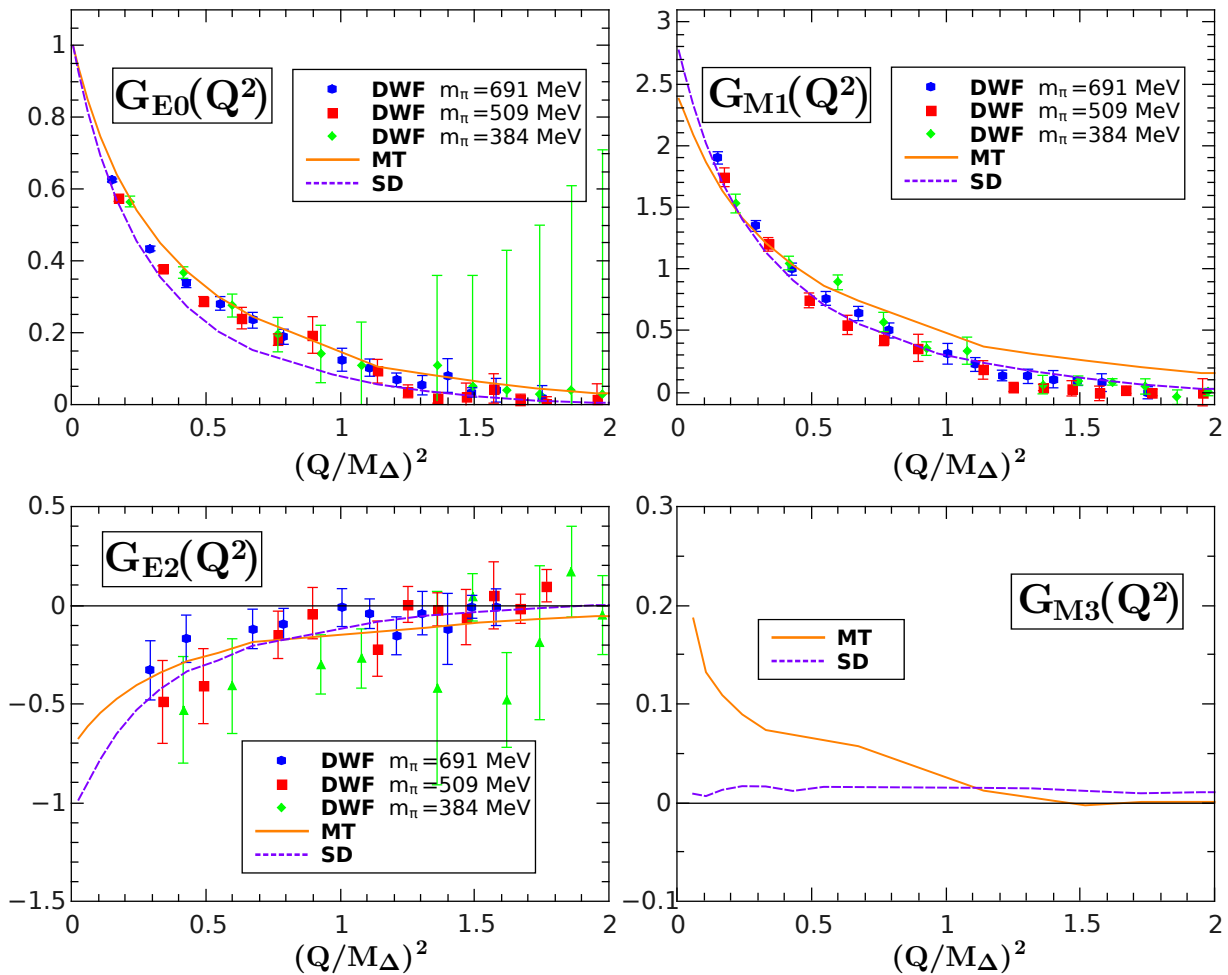}
 \end{center}
 \caption{Electromagnetic form factors for the $\Delta^{+}$ using the
Maris-tandy (MT) and the soft-divergence (SD) models. We compare with
unquenched lattice data (DWF) at three different pion masses
\cite{Alexandrou:2009hs,Alexandrou:2009nj}}\label{fig:DeltaFFs}
\end{figure*}

\begin{figure*}[btp]
 \begin{center}
  \includegraphics[width=0.9\textwidth,clip]{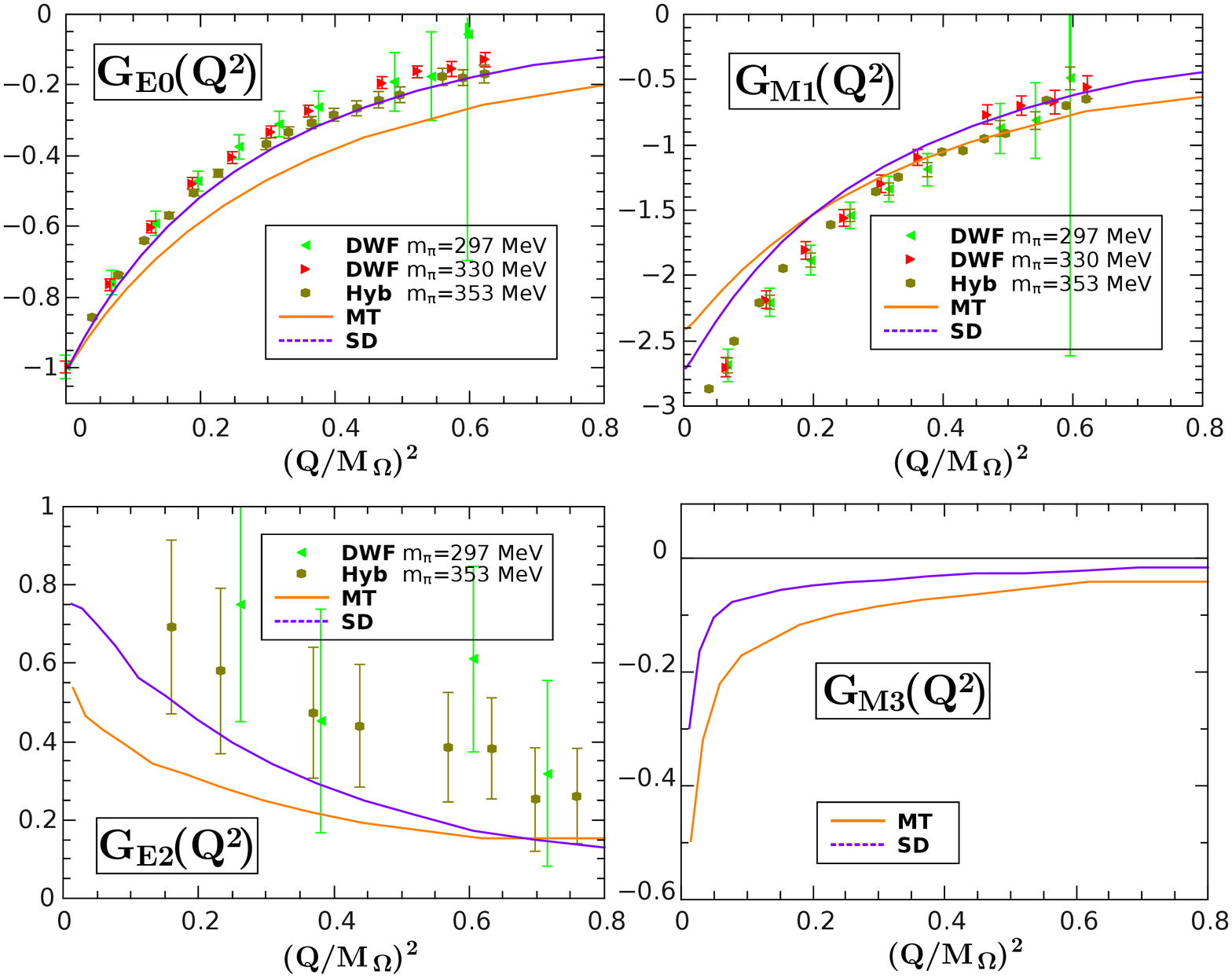}
 \end{center}
 \caption{Electromagnetic form factors for the $\Omega^{-}$ using the
Maris-tandy (MT) and the soft-divergence (SD) models. We compare with
unquenched (DWF) and mixed (Hyb.) lattice data at three different pion masses
\cite{Alexandrou:2010jv}}\label{fig:OmegaFFs}
\end{figure*}

%%%%%%%%%%%%%%%%%%%%%%%%%%%%%%%%%%%%%%%%%%%%%%%%%%%%%%%%%%%%%%%%%%%%%
%                      S U B S E C T I O N                          %
%%%%%%%%%%%%%%%%%%%%%%%%%%%%%%%%%%%%%%%%%%%%%%%%%%%%%%%%%%%%%%%%%%%%%

\subsection{Electric monopole form factor and charge radius}

The calculated electric monopole form factor $G_{E0}(Q^2)$ for the $\Delta^+$ is
shown in the upper-left panel of Fig. \ref{fig:DeltaFFs} and compared to lattice
calculations using dynamical Wilson fermions at three different pion masses
\cite{Alexandrou:2009hs,Alexandrou:2009nj}. The natural scale associated to the
problem is the Delta mass; since MT and SD models, as well as lattice
calculations, give different values for this mass, we plot the evolution of the
form factors in terms of the dimensionless quantity $Q^{2}/M^{2}$ to remove the
scale ambiguity that appears in the comparison of results using different
approaches/models. We stress again that, since we assume isospin
symmetry, the form factors for the $\Delta^{++}$, $\Delta^{0}$ and $\Delta^{-}$
are obtained by multiplying the former by the corresponding charge. 

We see from Fig. \ref{fig:DeltaFFs} that both the MT and the SD models show
good agreement with lattice calculations. The $Q^{2}$-evolution of  $G_{E0}$
differs slightly for the two models we considered. However, one must bear in
mind that we are working here with the simplest chiral-symmetry-preserving
interaction kernel (namely, the RL kernel). Since the effective couplings
 are tailored to
reproduce meson observables, we consider it sufficient if they reproduce baryon
properties at the level of a few percent. From this point of view, we can say that the
behavior of $G_{E0}(Q^2)$ is qualitatively model independent in our approach.

The  charge radius is calculated using the equation
\begin{equation}
 <r_{E0}^2>=-\frac{6}{G_{E0}(0)}\frac{dG_{E0}(Q^2)}{dQ^2}
\end{equation}
and the results are shown in Table \ref{tab:delta_moments} for the MT and SD
models as well as for lattice calculations. As before, we can suppress the scale
dependence of the charge radius by calculating the dimensionless quantity
$<r_{E0}^2>M_{\Delta}^2$. This quantity shows a better agreement with the
lattice data than the dimensionful charge radius does, although the value for the 
SD model is significantly larger. 

It is worth mentioning that chiral perturbation theory shows that, when the
$\Delta\to N\pi$ decay channel opens, the charge radius changes abruptly to a
lower value \cite{Ledwig:2011cx}. Since in our calculation we do not provide a
mechanism for the Delta to decay, it is therefore reasonable that in a full
calculation this would lead to a lower result for ${<r_{E0}^2>}$. This effect,
nevertheless, would be compensated partly by the inclusion of mesonic
effects.

\begin{table*}
\begin{center}
\small
\renewcommand{\arraystretch}{1.2}
\begin{tabular}{c|cc|ccc|c|} 
~ & F-MT & F-SD & DW1 & DW2 & DW3 & Exp.\\
\hline\hline
$M_\Delta (\textnormal{GeV})$ & 1.22 & 1.22 & 1.395~(18) & 1.559~(19)
& 1.687~(15) & 1.232~(2)
\\
\hline
$<r_{E0}^2> (\textnormal{fm}^2)$ & 0.50 & 0.61 & 0.373~(21) &
0.353~(12) & 0.279~(6) &
\\
\hline
$<r_{E0}^2> M_{\Delta}^2$ & 0.75 & 0.91 & 0.726~(36) & 0.858~(25) &
0.794~(14) &
\\
\hline
$G_{M1}(0)$ & 2.38 & 2.77 & 2.35~(16) & 2.68~(13) & 2.589~(78) &
3.54$^{+4.59}_{-4.72}$ \\ \hline
\end{tabular}
\caption{Comparison of results for the $\Delta^{+}$ mass, charge radius
$<r_{E0}^2>$ and for $G_{M1}(0)~~(\propto\mu)$. We compare our results for the
MT model (F-MT) and for the SD model (F-SD) with a lattice calculation with
dynamical Wilson fermions at $m_\pi=384$~MeV (DW1), $m_\pi=509$~MeV (DW2) and
$m_\pi=691$~MeV (DW3) \cite{Alexandrou:2009hs,Alexandrou:2009nj}. For
$G_{M1}(0)$ we also compare with the experimental value
\cite{Kotulla:2002cg,Beringer:1900zz}.}\label{tab:delta_moments}
\end{center}
\end{table*}

Since we asume isospin symmetry, in our framework the $\Omega$ baryon is
identical to the $\Delta$ but evaluated at a different current-quark mass. We
show the evolution of the electromagnetic form factors for the $\Omega^{-}$ in
Fig. \ref{fig:OmegaFFs}. The calculation shows good agreement with lattice data
for both models and, as before, a qualitative agreement between them. The
electric charge radius is shown in Table \ref{tab:omega_moments}. In this case
the calculated charge radius is smaller than the lattice values. However, the
dimensionless quantity $<r_{E0}^2>M_{\Omega}^2$ shows good agreement between our
results and lattice. Also, our result for this quantity shows little quark-mass
dependence, as can be seen by comparing the values for the $\Omega$ and the
$\Delta$; presumably, the inclusion of pion-cloud effects, or indeed other 
flavor dependent contributions
beyond that of Rainbow-Ladder, would account for the quark-mass dependence of 
the charge radius. 

\begin{table*}
\begin{center}
\small
\renewcommand{\arraystretch}{1.2}
\begin{tabular}{c|cc|ccc|c|}
~ & F-MT & F-SD & DW1 & DW2 & Hyb. & Exp.\\
\hline\hline
$M_\Omega (\textnormal{GeV})$ & 1.65 & 1.80 & 1.76~(2) & 1.77~(3) &
1.78~(3) & 1.672
\\
\hline
$<r_{E0}^2> (\textnormal{fm}^2)$ & 0.27 & 0.27 & 0.355~(14) &
0.353~(8) & 0.338~(9) &
\\
\hline
$<r_{E0}^2> M_{\Omega}^2$ & 0.74 & 0.89 & 0.726~(36) & 0.858~(25) &
0.794~(14) &
\\
\hline
$G_{M1}(0)$ & -2.41 & -2.71 & -3.443~(173) & -3.601~(109) & -3.368~(80) &
-3.52~(9) \\ \hline
\end{tabular}
\caption{Comparison of results for the $\Omega^{-}$ mass, charge radius
$<r_{E0}^2>$ and for $G_{M1}(0)~~(\propto\mu)$. We compare our results for the
MT model (F-MT) and for the SD model (F-SD) with a lattice calculation with
dynamical Wilson fermions at $m_\pi=297$~MeV (DW1), $m_\pi=330$~MeV (DW2) and
with a hybrid action at $m_\pi=353$~MeV (Hyb) \cite{Alexandrou:2010jv}. For
$G_{M1}(0)$ we also compare to the experimental value
\cite{Beringer:1900zz}.}\label{tab:omega_moments}
\end{center}
\end{table*}

%%%%%%%%%%%%%%%%%%%%%%%%%%%%%%%%%%%%%%%%%%%%%%%%%%%%%%%%%%%%%%%%%%%%%
%                      S U B S E C T I O N                          %
%%%%%%%%%%%%%%%%%%%%%%%%%%%%%%%%%%%%%%%%%%%%%%%%%%%%%%%%%%%%%%%%%%%%%

\subsection{Magnetic dipole form factor}

As already mentioned above, the magnetic moments of the $\Delta^{+}$ and
$\Delta^{++}$ are two of the few electromagnetic properties of the Delta for
which we have experimental input. The value at $Q^2=0$ of the magnetic dipole
form factor $G_{M1}(0)$ for the $\Delta^{+}$, which is related to the magnetic
moment via the relation
\begin{equation}\label{eq:magneticmoment}
 \mu_\Delta=\frac{e}{2M_\Delta}G_{M1}(0) \,\,,
\end{equation}
is given in Table \ref{tab:delta_moments}. We find good agreement between our
results and the lattice data at different pion masses. The value of $G_{M1}(0)$
for the $\Omega^{-}$ is shown in Table \ref{tab:omega_moments}. Here the
comparison with lattice is less favorable, and we clearly underestimate the
experimental value which, in this case, is very accurately measured. This is a
signature of missing meson-cloud effects whose relevance is, as discussed in the last section,
 somewhat obscured at
 the $u/d$ quark mass region since the effective interactions are fitted for that sector, thus
 in a sense incorporating pion-cloud  effects in the parameters of the model.

The evolution of $G_{M1}$ with the photon momentum also compares favorably with
lattice results in the case of the $\Delta$. Again, this is not the case for the
$\Omega$ as now both models differ significantly from lattice calculations at
low-$Q^2$, where pion- and kaon-cloud effects are expected to be more relevant.

%%%%%%%%%%%%%%%%%%%%%%%%%%%%%%%%%%%%%%%%%%%%%%%%%%%%%%%%%%%%%%%%%%%%%
%                      S U B S E C T I O N                          %
%%%%%%%%%%%%%%%%%%%%%%%%%%%%%%%%%%%%%%%%%%%%%%%%%%%%%%%%%%%%%%%%%%%%%

\subsection{Electric quadrupole form factor}

A non-vanishing value for the electric quadrupole moment signals the deformation
of the electric charge distribution from sphericity. It would be identically
zero if the baryon were formed only by s-wave components. In our approach, the
presence of higher angular-momentum components is a natural consequence of
requiring Poincar\'e covariance \cite{SanchisAlepuz:2011jn}. Nevertheless, the
relative importance of these components is dictated by the dynamics and we could
still obtain a non-trivial vanishing value for this moment.

We show our calculations for the electric quadrupole form factor and its
evolution with $Q^2$ in the bottom-right panel of Fig. \ref{fig:DeltaFFs}.
Although the precise vale of $G_{E2}(0)$ is very sensitive to numerical accuracy
(due to the presence of an $1/Q^4$ factor when extracting the form factor from
the electromagnetic current; see \Eq{eq:expressionGs}), we clearly see that
for both the MT and SD models it is non-vanishing and negative. In the
Breit frame (and for positively charged baryons), a negative value of
the electric quadrupole moment can be interpreted as an oblate distribution of
electric charge. This result agrees with lattice estimations, albeit in this
case lattice gives very noisy results and only for relatively high $Q$-values.

As expected, we obtain similar results for the $\Omega^{-}$, although with a
different sign coming from the $\Omega$ charge. The electric quadrupole form
factor is non-vanishing and negative, and therefore the charge distribution in
this case also features an oblate shape. This result agrees as well with the
available lattice data.

%%%%%%%%%%%%%%%%%%%%%%%%%%%%%%%%%%%%%%%%%%%%%%%%%%%%%%%%%%%%%%%%%%%%%
%                      S U B S E C T I O N                          %
%%%%%%%%%%%%%%%%%%%%%%%%%%%%%%%%%%%%%%%%%%%%%%%%%%%%%%%%%%%%%%%%%%%%%

\subsection{Magnetic octupole form factor}

Similar to the electric quadrupole moment, in the Breit frame the magnetic octupole moment measures
the deviation from sphericity of the magnetic dipole distribution. 

In the case of the magnetic octupole, we have to face an $1/Q^6$ factor when extracting the form factor from
the electromagnetic current. This entails that, with our current accuracy, we cannot give a reliable value for $G_{M3}(0)$, as is clearly
seen in the bottom-right panels of Figs. \ref{fig:DeltaFFs} and
\ref{fig:OmegaFFs}. However, in both cases and for both the MT and the SD models, we
can unambiguously say that the magnetic octupole moment is non-vanishing but small, and positive (negative for
the $\Omega^{-}$). We therefore predict a prolate distribution of the magnetic dipole.
Unfortunately, for the magnetic octupole form factor there are no reliable
lattice calculations to compare with, although a quenched calculation
\cite{Boinepalli:2009sq} suggests a negative sign for the $\Delta^{+}$, in contradiction to our findings. 
It is very well possible that a more elaborate truncation would change the sign of our results. However, it is
for us very difficult to estimate \textit{a priori} how the inclusion of, for instance, a pion-, resp., 
kaon-cloud would modify them.

%%%%%%%%%%%%%%%%%%%%%%%%%%%%%%%%%%%%%%%%%%%%%%%%%%%%%%%%%%%%%%%%%%%%%
%                        S E C T I O N                              %
%%%%%%%%%%%%%%%%%%%%%%%%%%%%%%%%%%%%%%%%%%%%%%%%%%%%%%%%%%%%%%%%%%%%%

\section{Summary}\label{sec:summary}

We have shown the calculation of the electromagnetic form factors of the
$\Delta$ and $\Omega$ baryons in the Poincar\'e-covariant BSE and DSE framework.
This framework has as a goal to provide an unified and systematically improvable
approach to hadron physics from continuum QCD. The calculation presented here
uses the Rainbow-Ladder truncation of the complete interaction kernel and within
this truncation scheme we solved self-consistently for all the elements in the
equations, namely the full quark propagator and quark-photon vertex. We have
performed the calculations using two different models for the dressings required
in the RL truncation, as an attempt to provide results which are qualitatively
model-independent.

Our results at $u/d$ quark mass show good agreement with lattice calculations
and are compatible with the few experimental data available for the $\Delta$. We
obtain a negative value of the electric quadrupole moment, indicating an oblate
charge distribution. The sign of the magnetic octupole moment is, however,
positive, which would correspond to a prolate magnetic dipole distribution. In
the absence of a proper treatment of the current-quark mass dependence of
mesonic effects or the quark-gluon interaction in our calculations, we 
find a weak dependence of the
electromagnetic properties on the current-quark mass. It is, therefore,
reasonable that we observe discrepancies between our results and lattice
calculations for the $\Omega$ form factors. 

This calculation, and especially the magnetic octupole form factor, is very
sensitive to numerical artefacts and for this reason the inaccuracy of the
results is sometimes significant. Improvements on our algorithms and the employment of more elaborate 
interaction kernels are thus
desirable in order to verify, in particular, the sign of the magnetic octupole moment.

%%%%%%%%%%%%%%%%%%%%%%%%%%%%%%%%%%%%%%%%%%%%%%%%%%%%%%%%%%%%%%%%%%%%%
%                 A C K N O W L E D G M E N T S                     %
%%%%%%%%%%%%%%%%%%%%%%%%%%%%%%%%%%%%%%%%%%%%%%%%%%%%%%%%%%%%%%%%%%%%%

\acknowledgments
We thank Gernot Eichmann, Christian S.\ Fischer and Selym Villalba-Chavez for
helpful discussions.

This work has been funded by the Austrian Science Fund, FWF, under project
P20592-N16. HSA acknowledges support by the Doctoral Program on Hadrons in Vacuum, 
Nuclei, and Stars (FWF DK W1203-N16) and funding by DFG through the TR16 project; RW acknowledges funding by the FWF 
under project M1333-N16. Further support by the  European Union (HadronPhysics2 
project ``Study of strongly-interacting matter'') is acknowledged.

\appendix

%%%%%%%%%%%%%%%%%%%%%%%%%%%%%%%%%%%%%%%%%%%%%%%%%%%%%%%%%%%%%%%%%%%%%
%                        S E C T I O N                              %
%%%%%%%%%%%%%%%%%%%%%%%%%%%%%%%%%%%%%%%%%%%%%%%%%%%%%%%%%%%%%%%%%%%%%

\section{Extraction of the form factors}\label{sec:appendix_current}

In Section \ref{sec:current} we derived an expression for the electromagnetic
current in terms of the photon interaction with the quarks forming a baryon. On
the other hand, the form of the current is constrained by Lorentz invariance and
current conservation to be a linear combination of a finite numbers of Lorentz
covariants with scalar coefficients. These coefficients are the form factors.

The electromagnetic current for a spin-$\nicefrac{3}{2}$ particle is
characterized by four form factors $F_i(Q^2)$
\cite{Nozawa:1990gt,Nicmorus:2010sd}. Its expression reads
\begin{align}
 J^{\mu,\alpha\beta}(P,Q)={}&\mathbb{P}^{\alpha\alpha'}(P_f)\left[
\left((F_1+F_2)i\gamma^\mu-F_2\frac{P^\mu}{M}\right)\delta^{\alpha'\beta'}
\right.\nonumber \\                         
&+\left.\left((F_3+F_4)i\gamma^\mu-F_4\frac{P^\mu}{M}\right)\frac{Q^{\alpha'}Q^{
\beta'}}{4M^2}\right]
\nonumber\\
& ~ \hspace{39mm} \mathbb{P}^{\beta'\beta}(P_i)
\end{align}
where $\mathbb{P}$ is the Rarita-Schwinger projector 
\begin{align}\label{eq:def_projectors}
\Lambda^+(\hat{P})=&\frac{1}{2}\left(\mathbb{1}+\Slash{\hat{P}}\right)~,\\
 \mathbb{P}_+^{\mu\nu}(\hat{\textnormal{P}})=&~\Lambda_+(\hat{\textnormal{P}})
 \left(T_P^{\mu\nu}-\frac{1}{3}\gamma^\mu_T\gamma^\nu_T\right)~,
\end{align}
with $\gamma^\mu_T=T_P^{\mu\nu}\gamma^\nu$, $T_P^{\mu\nu}$ the transverse
projector \Eq{eq:def_transverse_proj} and the hat denotes a unit vector.
$P_i$ and $P_f$ are the initial and final baryon total momenta, respectively, 
$Q=P_f-P_i$ is the photon momentum, $M$ is the baryon mass and $P=(P_f+P_i)/2$.
The form factors that are measured experimentally are the electric monopole
($G_{E_0}(Q^2)$), magnetic dipole ($G_{M_1}(Q^2)$), electric quadrupole
($G_{E_2}(Q^2)$) and magnetic octupole ($G_{M_3}(Q^2)$) form factors. They are
related to the $F_i's$ via \cite{Nozawa:1990gt}
\begin{align}
G_{E_0} &= \left(1+\frac{2\tau}{3}\right) ( F_1 - \tau F_2)  - \frac{\tau}{3}
(1+\tau) \,( F_3 - \tau F_4) \,, \\
G_{M_1} &= \left(1+\frac{4\tau}{5}\right) (F_1+F_2) - \frac{2\tau}{5} (1+\tau)\,
(F_3 + F_4)\,,\\
G_{E_2} &= (F_1 - \tau F_2) - \frac{1}{2}\,(1+\tau) \,(F_3 - \tau F_4)\,,\\
G_{M_3} &= (F_1 + F_2) - \frac{1}{2}\,(1+\tau) \,(F_3 + F_4)\,,
\end{align} 
with $\tau=Q^2/4M^2$. It is shown in \cite{Nozawa:1990gt} that if charge and
magnetic dipole distribution in the baryon is spherically symmetric then
$G_{E_2}$ and $G_{M_3}$ must vanish, respectively; therefore they measure the
deformation of the object. At $Q^2=0$ the form factors define the electric
charge ($e_{\nicefrac{3}{2}}$), magnetic dipole moment
($\mu_{\nicefrac{3}{2}}$), electric quadrupole moment
($\mathcal{Q}_{\nicefrac{3}{2}}$) and magnetic octupole moment
($\mathcal{O}_{\nicefrac{3}{2}}$) of a spin-$\nicefrac{3}{2}$ particle,
\begin{align}
e_{\nicefrac{3}{2}}&=G_{E_0}(0)\,, \\
\mu_{\nicefrac{3}{2}}&=\frac{e}{2M}G_{M_1}(0)\,,\\
\mathcal{Q}_{\nicefrac{3}{2}}&=\frac{e}{M^2}G_{E_2}(0)\,,\\
\mathcal{O}_{\nicefrac{3}{2}}&=\frac{e}{2M^3}G_{M_3}(0)\,.
\end{align} 

Once the electromagnetic current is calculated from (\ref{eq:FFeqRL}), the form
factors can be extracted using the expressions \cite{Nicmorus:2010sd}
\begin{align}\label{eq:expressionGs}
G_{E_0} &= \frac{s_2-2s_1}{4i\sqrt{1+\tau}}\,, \\
G_{M_1} &= \frac{9i}{40\,\tau}\left(s_4-2s_3\right)\,, \\
G_{E_2} &= \frac{3}{8i\,\tau^2\sqrt{1+\tau}} \left[ 2s_1
\left(\tau+\frac{3}{2}\right) - \tau s_2 \right], \\
G_{M_3} &= \frac{3i}{16\,\tau^3} \left[ 2s_3 \left(\tau+\frac{5}{4}\right) -
\tau s_4 \right]~,
\end{align} 
where

\begin{align}\label{sses}
s_1 &= \textnormal{Tr}\left\{ J^{\mu,\alpha\beta}  \hat{P}^\mu  \hat{P}^\alpha 
\hat{P}^\beta \right\}~, \\
s_2 &= \textnormal{Tr}\left\{ J^{\mu,\alpha\alpha}  \hat{P}^\mu \right\}~, \\
s_3 &= \textnormal{Tr}\left\{ J^{\mu,\alpha\beta} \,\gamma^\mu_T  \hat{P}^\alpha
 \hat{P}^\beta \right\}~, \\
s_4 &= \textnormal{Tr}\left\{ J^{\mu,\alpha\alpha} \,\gamma^\mu_T \right\} \,.
\end{align} 


\begin{thebibliography}{99}


%\cite{Nefkens:1977eb}
\bibitem{Nefkens:1977eb}
  B.~M.~K.~Nefkens, M.~Arman, H.~C.~Ballagh, Jr., P.~F.~Glodis, R.~P.~Haddock, K.~C.~Leung, D.~E.~A.~Smith and D.~I.~Sober,
  %``Differential Cross-Sections for Pion-Proton Bremsstrahlung at 269-MeV, 298-MeV, and 324-MeV,''
  Phys.\ Rev.\ D {\bf 18} (1978) 3911.
  %%CITATION = PHRVA,D18,3911;%%

%\cite{Heller:1986gn}
\bibitem{Heller:1986gn}
  L.~Heller, S.~Kumano, J.~C.~Martinez and E.~J.~Moniz,
  %``Pion - Nucleon Bremsstrahlung And Delta Electromagnetic Moments,''
  Phys.\ Rev.\ C {\bf 35} (1987) 718.
  %%CITATION = PHRVA,C35,718;%%

%\cite{Wittman:1987kb}
\bibitem{Wittman:1987kb}
  R.~Wittman,
  %``A COVARIANT APPROACH TO pi+ PROTON BREMSSTRAHLUNG,''
  Phys.\ Rev.\ C {\bf 37} (1988) 2075.
  %%CITATION = PHRVA,C37,2075;%%

%\cite{Lin:1991tx}
\bibitem{Lin:1991tx}
  D.~Lin and M.~K.~Liou,
  %``Soft photon analysis of pion - proton bremsstrahlung and the 'experimental' magnetic moment of Delta (1232)++,''
  Phys.\ Rev.\ C {\bf 43} (1991) 930.
  %%CITATION = PHRVA,C43,930;%%

%\cite{Lin:1991qk}
\bibitem{Lin:1991qk}
  D.~H.~Lin, M.~K.~Liou and Z.~M.~Ding,
  %``Pion proton bremsstrahlung calculation and the 'experimental' magnetic
%moment of Delta++ (1232),''
  Phys.\ Rev.\ C {\bf 44} (1991) 1819.
  %%CITATION = PHRVA,C44,1819;%%

%\cite{Bosshard:1990ys}
\bibitem{Bosshard:1990ys}
  A.~Bosshard, C.~Amsler, J.~A.~Bistirlich, B.~van den Brandt, K.~M.~Crowe, M.~Doebeli, M.~Doser and R.~P.~Haddock {\it et al.},
  %``Polarized target asymmetry in pion proton bremsstrahlung at 298-MeV,''
  Phys.\ Rev.\ Lett.\  {\bf 64} (1990) 2619.
  %%CITATION = PRLTA,64,2619;%%

%\cite{Bosshard:1991zp}
\bibitem{Bosshard:1991zp}
  A.~Bosshard, C.~Amsler, M.~Doebeli, M.~Doser, M.~Schaad, J.~Riedlberger, P.~Truoel and J.~A.~Bistirlich {\it et al.},
  %``Analyzing power in pion proton bremsstrahlung, and the Delta++ (1232) magnetic moment,''
  Phys.\ Rev.\ D {\bf 44} (1991) 1962.
  %%CITATION = PHRVA,D44,1962;%%

%\cite{LopezCastro:2000cv}
\bibitem{LopezCastro:2000cv}
  G.~Lopez Castro and A.~Mariano,
  %``Determination of the Delta++ magnetic dipole moment,''
  Phys.\ Lett.\ B {\bf 517} (2001) 339
  [nucl-th/0006031].
  %%CITATION = NUCL-TH/0006031;%%

%\cite{LopezCastro:2000ep}
\bibitem{LopezCastro:2000ep}
  G.~Lopez Castro and A.~Mariano,
  %``Elastic and radiative pi+ p scattering and properties of the Delta++ resonance,''
  Nucl.\ Phys.\ A {\bf 697} (2002) 440
  [nucl-th/0010045].
  %%CITATION = NUCL-TH/0010045;%%

%\cite{Beringer:1900zz}
\bibitem{Beringer:1900zz}
  J.~Beringer {\it et al.}  [Particle Data Group Collaboration],
  %``Review of Particle Physics (RPP),''
  Phys.\ Rev.\ D {\bf 86} (2012) 010001.
  %%CITATION = PHRVA,D86,010001;%%

%\cite{Kotulla:2002cg}
\bibitem{Kotulla:2002cg}
  M.~Kotulla, J.~Ahrens, J.~R.~M.~Annand, R.~Beck, G.~Caselotti, L.~S.~Fog, D.~Hornidge and S.~Janssen {\it et al.},
  %``The Reaction gamma p ---> pi zero gamma-prime p and the magnetic dipole moment of the delta+(1232) resonance,''
  Phys.\ Rev.\ Lett.\  {\bf 89} (2002) 272001
  [nucl-ex/0210040].
  %%CITATION = NUCL-EX/0210040;%%

%\cite{Blanpied:2001ae}
\bibitem{Blanpied:2001ae}
  G.~Blanpied, M.~Blecher, A.~Caracappa, R.~Deininger, C.~Djalali, G.~Giordano, K.~Hicks and S.~Hoblit {\it et al.},
  %``N ---> delta transition and proton polarizabilities from measurements of p (gamma polarized, gamma), p (gamma polarized, pi0), and p (gamma polarized, pi+),''
  Phys.\ Rev.\ C {\bf 64} (2001) 025203.
  %%CITATION = PHRVA,C64,025203;%%

%\cite{Alexandrou:2009hs}
\bibitem{Alexandrou:2009hs}
  C.~Alexandrou, T.~Korzec, G.~Koutsou, C.~Lorce, J.~W.~Negele, V.~Pascalutsa, A.~Tsapalis and M.~Vanderhaeghen,
  %``Quark transverse charge densities in the Delta(1232) from lattice QCD,''
  Nucl.\ Phys.\ A {\bf 825} (2009) 115
  [arXiv:0901.3457 [hep-lat]].
  %%CITATION = ARXIV:0901.3457;%%

%\cite{Alexandrou:2009nj}
\bibitem{Alexandrou:2009nj}
  C.~Alexandrou, T.~Korzec, G.~Koutsou, C.~Lorce, V.~Pascalutsa, M.~Vanderhaeghen, J.~W.~Negele and A.~Tsapalis,
  %``Delta electromagnetic form factors and quark transverse charge densities from lattice QCD,''
  PoS CD {\bf 09} (2009) 092
  [arXiv:0910.3315 [hep-lat]].
  %%CITATION = ARXIV:0910.3315;%%

%\cite{Alexandrou:2010jv}
\bibitem{Alexandrou:2010jv}
  C.~Alexandrou, T.~Korzec, G.~Koutsou, J.~W.~Negele and Y.~Proestos,
  %``The Electromagnetic form factors of the $\Omega^-$ in lattice QCD,''
  Phys.\ Rev.\ D {\bf 82} (2010) 034504
  [arXiv:1006.0558 [hep-lat]].
  %%CITATION = ARXIV:1006.0558;%%

%\cite{Aubin:2008qp}
\bibitem{Aubin:2008qp}
  C.~Aubin, K.~Orginos, V.~Pascalutsa and M.~Vanderhaeghen,
  %``Magnetic Moments of Delta and Omega- Baryons with Dynamical Clover Fermions,''
  Phys.\ Rev.\ D {\bf 79} (2009) 051502
  [arXiv:0811.2440 [hep-lat]].
  %%CITATION = ARXIV:0811.2440;%%

%\cite{Boinepalli:2009sq}
\bibitem{Boinepalli:2009sq}
  S.~Boinepalli, D.~B.~Leinweber, P.~J.~Moran, A.~G.~Williams, J.~M.~Zanotti and J.~B.~Zhang,
  %``Precision electromagnetic structure of decuplet baryons in the chiral regime,''
  Phys.\ Rev.\ D {\bf 80} (2009) 054505
  [arXiv:0902.4046 [hep-lat]].
  %%CITATION = ARXIV:0902.4046;%%

%\cite{Ramalho:2008dc}
\bibitem{Ramalho:2008dc}
  G.~Ramalho and M.~T.~Pena,
  %``Electromagnetic form factors of the Delta in a S-wave approach,''
  J.\ Phys.\ G G {\bf 36} (2009) 085004
  [arXiv:0807.2922 [hep-ph]].
  %%CITATION = ARXIV:0807.2922;%%

%\cite{Ramalho:2009vc}
\bibitem{Ramalho:2009vc}
  G.~Ramalho, M.~T.~Pena and F.~Gross,
  %``Electric quadrupole and magnetic octupole moments of the Delta,''
  Phys.\ Lett.\ B {\bf 678} (2009) 355
  [arXiv:0902.4212 [hep-ph]].
  %%CITATION = ARXIV:0902.4212;%%

%\cite{Ramalho:2010xj}
\bibitem{Ramalho:2010xj}
  G.~Ramalho, M.~T.~Pena and F.~Gross,
  %``Electromagnetic form factors of the Delta with D-waves,''
  Phys.\ Rev.\ D {\bf 81} (2010) 113011
  [arXiv:1002.4170 [hep-ph]].
  %%CITATION = ARXIV:1002.4170;%%

%\cite{Ramalho:2010rr}
\bibitem{Ramalho:2010rr}
  G.~Ramalho and M.~T.~Pena,
  %``Extracting the Omega- electric quadrupole moment from lattice QCD data,''
  Phys.\ Rev.\ D {\bf 83} (2011) 054011
  [arXiv:1012.2168 [hep-ph]].
  %%CITATION = ARXIV:1012.2168;%%

%\cite{Ledwig:2008es}
\bibitem{Ledwig:2008es}
  T.~Ledwig, A.~Silva and M.~Vanderhaeghen,
  %``Electromagnetic properties of the Delta(1232) and decuplet baryons in the self-consistent SU(3) chiral quark-soliton model,''
  Phys.\ Rev.\ D {\bf 79} (2009) 094025
  [arXiv:0811.3086 [hep-ph]].
  %%CITATION = ARXIV:0811.3086;%%

%\cite{Geng:2009ys}
\bibitem{Geng:2009ys}
  L.~S.~Geng, J.~Martin Camalich and M.~J.~Vicente Vacas,
  %``Electromagnetic structure of the lowest-lying decuplet resonances in covariant chiral perturbation theory,''
  Phys.\ Rev.\ D {\bf 80} (2009) 034027
  [arXiv:0907.0631 [hep-ph]].
  %%CITATION = ARXIV:0907.0631;%%

%\cite{Ledwig:2011cx}
\bibitem{Ledwig:2011cx}
  T.~Ledwig, J.~Martin-Camalich, V.~Pascalutsa and M.~Vanderhaeghen,
  %``The Nucleon and $\Delta$(1232) form factors at low momentum-transfer and small pion masses,''
  Phys.\ Rev.\ D {\bf 85} (2012) 034013
  [arXiv:1108.2523 [hep-ph]].
  %%CITATION = ARXIV:1108.2523;%%

%\cite{Azizi:2008tx}
\bibitem{Azizi:2008tx}
  K.~Azizi,
  %``Magnetic Dipole, Electric Quadrupole and Magnetic Octupole Moments of the Delta Baryons in Light Cone QCD Sum Rules,''
  Eur.\ Phys.\ J.\ C {\bf 61} (2009) 311
  [arXiv:0811.2670 [hep-ph]].
  %%CITATION = ARXIV:0811.2670;%%
  
  %\cite{SanchisAlepuz:2010in}
\bibitem{SanchisAlepuz:2010in}
  H.~Sanchis-Alepuz, R.~Alkofer, G.~Eichmann and S.~Villalba-Chavez,
  %``On baryon properties from a covariant Faddeev approach,''
  PoS LC {\bf 2010} (2010) 018
  [arXiv:1010.6183 [hep-ph]].
  %%CITATION = ARXIV:1010.6183;%%

%\cite{SanchisAlepuz:n}
\bibitem{SanchisAlepuz:2011jn}
  H.~Sanchis-Alepuz, G.~Eichmann, S.~Villalba-Chavez and R.~Alkofer,
  %``Delta and Omega masses in a three-quark covariant Faddeev approach,''
  Phys.\ Rev.\ D {\bf 84} (2011) 096003
  [arXiv:1109.0199 [hep-ph]].
  %%CITATION = ARXIV:1109.0199;%%

%\cite{SanchisAlepuz:2012vb}
\bibitem{SanchisAlepuz:2012vb}
  H.~Sanchis-Alepuz,
  %``Baryon properties and glueballs from Poincare-covariant bound-state equations,''
  arXiv:1206.5190 [hep-ph].
  %%CITATION = ARXIV:1206.5190;%%

%\cite{Haberzettl:1997jg}
\bibitem{Haberzettl:1997jg}
  H.~Haberzettl,
  %``Gauge invariant theory of pion photoproduction with dressed hadrons,''
  Phys.\ Rev.\ C {\bf 56} (1997) 2041
  [nucl-th/9704057].
  %%CITATION = NUCL-TH/9704057;%%

%\cite{Kvinikhidze:1998xn}
\bibitem{Kvinikhidze:1998xn}
  A.~N.~Kvinikhidze and B.~Blankleider,
  %``Gauging of equations method. 1. electromagnetic currents of three distinguishable particles,''
  Phys.\ Rev.\ C {\bf 60} (1999) 044003
  [nucl-th/9901001].
  %%CITATION = NUCL-TH/9901001;%%

%\cite{Kvinikhidze:1999xp}
\bibitem{Kvinikhidze:1999xp}
  A.~N.~Kvinikhidze and B.~Blankleider,
  %``Gauging of equations method. 2. Electromagnetic currents of three identical particles,''
  Phys.\ Rev.\ C {\bf 60} (1999) 044004
  [nucl-th/9901002].
  %%CITATION = NUCL-TH/9901002;%%

%\cite{Oettel:1999gc}
\bibitem{Oettel:1999gc}
  M.~Oettel, M.~Pichowsky and L.~von Smekal,
  %``Current conservation in the covariant quark diquark model of the nucleon,''
  Eur.\ Phys.\ J.\ A {\bf 8} (2000) 251
  [nucl-th/9909082];
  %%CITATION = NUCL-TH/9909082;%%
 %\cite{Oettel:2000jj}
%\bibitem{Oettel:2000jj}
  M.~Oettel, R.~Alkofer and L.~von Smekal,
  %``Nucleon properties in the covariant quark diquark model,''
  Eur.\ Phys.\ J.\ A {\bf 8} (2000) 553
  [nucl-th/0006082].
  %%CITATION = NUCL-TH/0006082;%%
  
%\cite{Munczek:1994zz}
\bibitem{Munczek:1994zz}
  H.~J.~Munczek,
  %``Dynamical chiral symmetry breaking, Goldstone's theorem and the consistency of the Schwinger-Dyson and Bethe-Salpeter Equations,''
  Phys.\ Rev.\ D {\bf 52} (1995) 4736
  [hep-th/9411239].
  %%CITATION = HEP-TH/9411239;%%

%\cite{Bender:1996bb}
\bibitem{Bender:1996bb}
  A.~Bender, C.~D.~Roberts and L.~Von Smekal,
  %``Goldstone theorem and diquark confinement beyond rainbow ladder approximation,''
  Phys.\ Lett.\ B {\bf 380} (1996) 7
  [nucl-th/9602012].
  %%CITATION = NUCL-TH/9602012;%%

%\cite{Maris:1997hd}
\bibitem{Maris:1997hd}
  P.~Maris, C.~D.~Roberts and P.~C.~Tandy,
  %``Pion mass and decay constant,''
  Phys.\ Lett.\ B {\bf 420} (1998) 267
  [nucl-th/9707003].
  %%CITATION = NUCL-TH/9707003;%%
  
%\cite{Bhagwat:2006pu}
\bibitem{Bhagwat:2006pu}
  M.~S.~Bhagwat and P.~Maris,
  %``Vector meson form factors and their quark-mass dependence,''
  Phys.\ Rev.\ C {\bf 77} (2008) 025203
  [nucl-th/0612069].
  %%CITATION = NUCL-TH/0612069;%%


%\cite{Maris:1997tm}
\bibitem{Maris:1997tm}
  P.~Maris and C.~D.~Roberts,
  %``Pi- and K meson Bethe-Salpeter amplitudes,''
  Phys.\ Rev.\ C {\bf 56} (1997) 3369
  [nucl-th/9708029].
  %%CITATION = NUCL-TH/9708029;%%

%\cite{Maris:1999nt}
\bibitem{Maris:1999nt}
  P.~Maris and P.~C.~Tandy,
  %``Bethe-Salpeter study of vector meson masses and decay constants,''
  Phys.\ Rev.\ C {\bf 60} (1999) 055214
  [nucl-th/9905056].
  %%CITATION = NUCL-TH/9905056;%%

%\cite{Alkofer:2008et}
\bibitem{Alkofer:2008et}
  R.~Alkofer, C.~S.~Fischer and R.~Williams,
  %``U(A)(1) anomaly and eta-prime mass from an infrared singular quark-gluon vertex,''
  Eur.\ Phys.\ J.\ A {\bf 38} (2008) 53
  [arXiv:0804.3478 [hep-ph]].
  %%CITATION = ARXIV:0804.3478;%%

%\cite{Alkofer:2008tt}
\bibitem{Alkofer:2008tt}
  R.~Alkofer, C.~S.~Fischer, F.~J.~Llanes-Estrada and K.~Schwenzer,
  %``The Quark-gluon vertex in Landau gauge QCD: Its role in dynamical chiral symmetry breaking and quark confinement,''
  Annals Phys.\  {\bf 324} (2009) 106
  [arXiv:0804.3042 [hep-ph]].
  %%CITATION = ARXIV:0804.3042;%%
  
%\cite{Fischer:2009jm}
\bibitem{Fischer:2009jm}
  C.~S.~Fischer and R.~Williams,
  %``Probing the gluon self-interaction in light mesons,''
  Phys.\ Rev.\ Lett.\  {\bf 103} (2009) 122001
  [arXiv:0905.2291 [hep-ph]].
  %%CITATION = ARXIV:0905.2291;%%
 
%\cite{Williams:2009wx}
\bibitem{Williams:2009wx}
  R.~Williams,
  %``Bethe-Salpeter studies of mesons beyond rainbow-ladder,''
  EPJ Web Conf.\  {\bf 3} (2010) 03005
  [arXiv:0912.3494 [hep-ph]].
  %%CITATION = ARXIV:0912.3494;%%
  
  %\cite{Chang:2009zb}  
  \bibitem{Chang:2009zb}
  L.~Chang and C.~D.~Roberts,
  %``Sketching the Bethe-Salpeter kernel,''
  Phys.\ Rev.\ Lett.\  {\bf 103} (2009) 081601
  [arXiv:0903.5461 [nucl-th]].
  %%CITATION = ARXIV:0903.5461;%%

%\cite{Fischer:2007ze}
\bibitem{Fischer:2007ze}
  C.~S.~Fischer, D.~Nickel and J.~Wambach,
  %``Hadronic unquenching effects in the quark propagator,''
  Phys.\ Rev.\ D {\bf 76} (2007) 094009
  [arXiv:0705.4407 [hep-ph]].
  %%CITATION = ARXIV:0705.4407;%%

%\cite{Fischer:2008wy}
\bibitem{Fischer:2008wy}
  C.~S.~Fischer and R.~Williams,
  %``Beyond the rainbow: Effects from pion back-coupling,''
  Phys.\ Rev.\ D {\bf 78} (2008) 074006
  [arXiv:0808.3372 [hep-ph]].
  %%CITATION = ARXIV:0808.3372;%%

  %\cite{SanchisAlepuz:2011aa}
\bibitem{SanchisAlepuz:2011aa}
  H.~Sanchis-Alepuz, R.~Alkofer, G.~Eichmann and R.~Williams,
  %``Model Comparison of Delta and Omega Masses in a Covariant Faddeev Approach,''
  PoS QCD{\bf-TNT-II} (2011) 041
  [arXiv:1112.3214 [hep-ph]].
  %%CITATION = ARXIV:1112.3214;%%

%\cite{Krassnigg:2009zh}
\bibitem{Krassnigg:2009zh}
  A.~Krassnigg,
  %``Survey of J=0,1 mesons in a Bethe-Salpeter approach,''
  Phys.\ Rev.\ D {\bf 80} (2009) 114010
  [arXiv:0909.4016 [hep-ph]].
  %%CITATION = ARXIV:0909.4016;%%

%\cite{Eichmann:2011vu}
\bibitem{Eichmann:2011vu}
  G.~Eichmann,
  %``Nucleon electromagnetic form factors from the covariant Faddeev equation,''
  Phys.\ Rev.\ D {\bf 84} (2011) 014014
  [arXiv:1104.4505 [hep-ph]].
  %%CITATION = ARXIV:1104.4505;%%

%\cite{Nicmorus:2010mc}
\bibitem{Nicmorus:2010mc}
  D.~Nicmorus, G.~Eichmann, A.~Krassnigg and R.~Alkofer,
  %``Delta properties in the rainbow-ladder truncation of Dyson-Schwinger equations,''
  Few Body Syst.\  {\bf 49} (2011) 255
  [arXiv:1008.4149 [hep-ph]].
  %%CITATION = ARXIV:1008.4149;%%

%\cite{Kogut:1973ab}
\bibitem{Kogut:1973ab}
  J.~B.~Kogut and L.~Susskind,
  %``Quark Confinement And The Puzzle Of The Ninth Axial Current,''
  Phys.\ Rev.\ D {\bf 10} (1974) 3468.
  %%CITATION = PHRVA,D10,3468;%%

%\cite{vonSmekal:1997dq}
\bibitem{vonSmekal:1997dq}
  L.~von Smekal, A.~Mecke and R.~Alkofer,
  %``A Dynamical eta-prime mass from an infrared enhanced gluon exchange,''
  In *Big Sky 1997, Intersections between particle and nuclear physics* 746-749
  [hep-ph/9707210].
  %%CITATION = HEP-PH/9707210;%%

%\cite{Alkofer:2003jk}
\bibitem{Alkofer:2003jk}
  R.~Alkofer, W.~Detmold, C.~S.~Fischer and P.~Maris,
  %``Analytic structure of the gluon and quark propagators in Landau gauge QCD,''
  Nucl.\ Phys.\ Proc.\ Suppl.\  {\bf 141} (2005) 122
  [hep-ph/0309078].
  %%CITATION = HEP-PH/0309078;%%

%\cite{Alkofer:2003jj}
\bibitem{Alkofer:2003jj}
  R.~Alkofer, W.~Detmold, C.~S.~Fischer and P.~Maris,
  %``Analytic properties of the Landau gauge gluon and quark propagators,''
  Phys.\ Rev.\ D {\bf 70} (2004) 014014
  [hep-ph/0309077].
  %%CITATION = HEP-PH/0309077;%%
  
%\cite{Thomas:2007bc}
\bibitem{Thomas:2007bc}
  A.~W.~Thomas,
  %``The Pion cloud: Insights into hadron structure,''
  Prog.\ Theor.\ Phys.\  {\bf 168} (2007) 614
  [arXiv:0711.2259 [nucl-th]].
  %%CITATION = ARXIV:0711.2259;%%
 
  %\cite{Eichmann:2008ae}
\bibitem{Eichmann:2008ae}
G.~Eichmann {\it et al.}, 
 % G.~Eichmann, R.~Alkofer, I.~C.~Cloet, A.~Krassnigg and C.~D.~Roberts,
  %``Perspective on rainbow-ladder truncation,''
  Phys.\ Rev.\ C {\bf 77} (2008) 042202
  [arXiv:0802.1948 [nucl-th]].
  %%CITATION = ARXIV:0802.1948;%%
  
%\cite{SanchisAlepuz:2012ej}
\bibitem{SanchisAlepuz:2012ej}
  H.~Sanchis-Alepuz, R.~Alkofer and R.~Williams,
  %``Baryon Properties from the Covariant Faddeev Equation,''
  PoS QNP {\bf 2012} (2012) 112
  [arXiv:1206.6599 [hep-ph]]; 
  %%CITATION = ARXIV:1206.6599;%%
PoS Confinement X (2013) 101.

%\cite{Nozawa:1990gt}
\bibitem{Nozawa:1990gt}
  S.~Nozawa and D.~B.~Leinweber,
  %``Electromagnetic form-factors of spin 3/2 baryons,''
  Phys.\ Rev.\ D {\bf 42} (1990) 3567.
  %%CITATION = PHRVA,D42,3567;%%

%\cite{Nicmorus:2010sd}
\bibitem{Nicmorus:2010sd}
  D.~Nicmorus, G.~Eichmann and R.~Alkofer,
  %``Delta and Omega electromagnetic form factors in a Dyson-Schwinger/Bethe-Salpeter approach,''
  Phys.\ Rev.\ D {\bf 82} (2010) 114017
  [arXiv:1008.3184 [hep-ph]].
  %%CITATION = ARXIV:1008.3184;%%

\end{thebibliography}
\end{document}